\documentclass[prl,preprint,longbibliography,floatfix]{revtex4-1}
\usepackage{color}
\usepackage{graphicx}
\usepackage{mathrsfs}
\usepackage{bm}

\begin{document}

\title{Topological states in engineered atomic lattices}

\author{Robert Drost}
\author{Teemu Ojanen}
\author{Ari Harju}
\author{Peter Liljeroth}
\email{peter.liljeroth@aalto.fi}
\affiliation{Department of Applied Physics, Aalto University School of Science, PO Box 15100, 00076 Aalto, Finland}

\begin{abstract}
	
\textbf{Topological materials exhibit protected edge modes that have been proposed for applications in for example spintronics and quantum computation. While a number of such systems exist \cite{Kane2005,Bernevig2006,Qi2011,Mourik2012,Nadj-Perge2014,Cheon2015}, it would be desirable to be able to test theoretical proposals in an artificial system that allows precise control over the key parameters of the model \cite{Gomes2012}. The essential physics of several topological systems can be captured by tight-binding models, which can also be implemented in artificial lattices \cite{Gomes2012,Cheon2015}. Here, we show that this method can be realized in a vacancy lattice in a chlorine monolayer on a Cu(100) surface. We use low-temperature scanning tunneling microscopy (STM) to fabricate such lattices with atomic precision and probe the resulting local density of states (LDOS) with scanning tunneling spectroscopy (STS). We create analogues of two tight-binding models of fundamental importance: The polyacetylene (dimer) chain with topological domain wall states, and the Lieb lattice, featuring a lattice pseudospin 1 system with a flat electron band. These results provide the first steps in realizing designer quantum materials with tailored properties.}
	
\end{abstract}
\maketitle

The topological theory of matter has its roots in the 70s and 80s studies of polyacetyle \cite{heeger}, the quantum Hall effect \cite{Klitzing1980} and superfluid $^3$He \cite{volovik}. The interest in the field exploded after the 2006 discovery of topological insulators \cite{Bernevig2006,Qi2011}. Crystalline solids exhibit a spectrum of energy bands constrained by the materials symmetries. Topological properties of the band structure give rise to protected boundary states that are the hallmark of topological materials. Well-known examples include chiral and helical states in quantum (spin) Hall systems and Majorana modes in topological superconductors \cite{Klitzing1980, Kane2005, konig2007, das2012, nadj2014, albrecht2016}. A central goal of research has become to identify systems with a topological spectrum.

Theories of topological materials \cite{Schnyder2008} are often formulated using tight-binding models describing hopping between localized electronic orbitals. This means that, given sufficient control, one can implement these models by assembling the corresponding structure from individual constituents that are suitably coupled. It is hence possible to build 'designer quantum materials' based on specific Hamiltonians through atomic assemblies. The STM is the central tool to fabricate and characterize structures at the atomic scale \cite{Crommie1993,Manoharan2000,Hirjibehedin2006, Gomes2012,Paavilainen2016}. Recently, it was shown that vacancy defects in the c($2\times2$) chlorine superstructure on Cu(100) make an excellent system for large scale atomic assembly \cite{Kalff2016}. We show that individual Cl vacancies host a well-defined vacancy state below the band edge of the chlorine layer which interact when sufficiently close to each other. We demonstrate that it is possible to construct coupled lattices and implemented two Hamiltonians of general interest, the Su-Schrieffer-Heeger (SSH) dimer chain and the 2D Lieb lattice.

\begin{figure}[h!]
	\centering
	\includegraphics[width=0.450\textwidth]{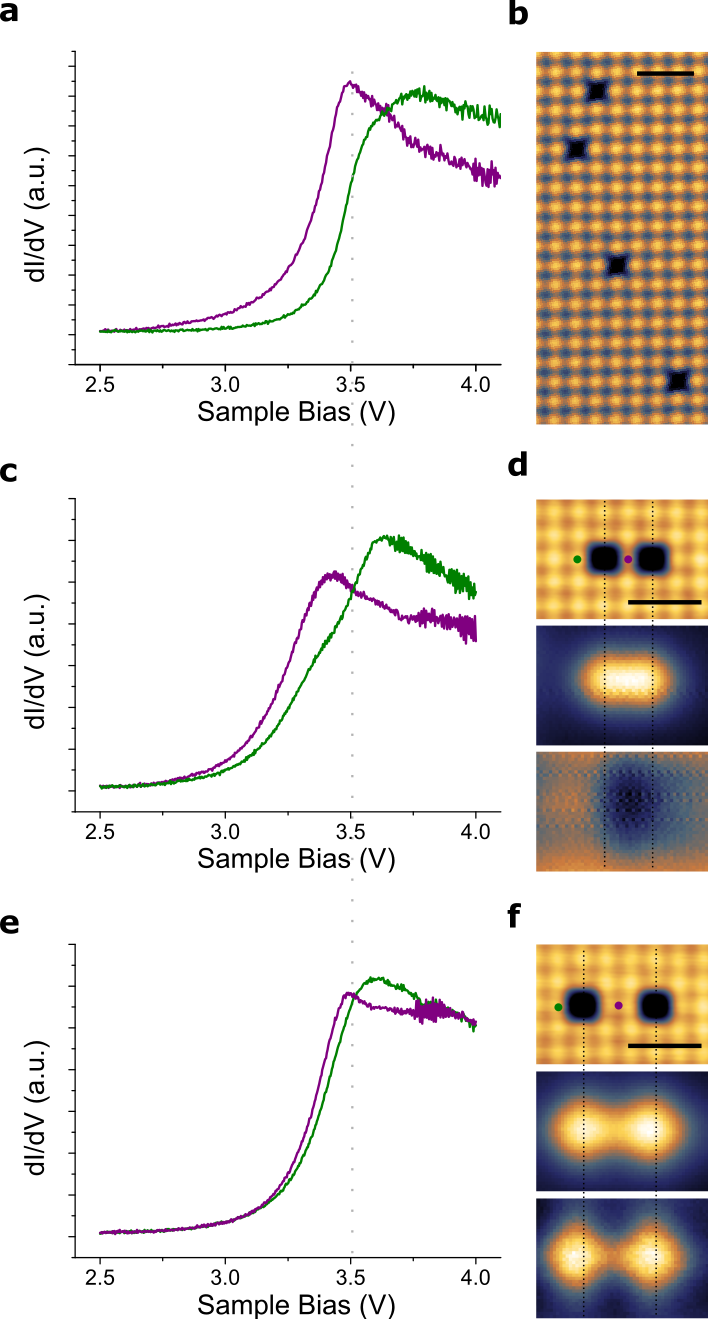}
	\caption{\textbf{Electronic structure of vacancies and vacancy dimers.} \textbf{a}, Typical conductance spectra acquired on the chlorine layer (green) and on and isolated vacancy (purple). \textbf{b}, Overview image of the chlorine terminated Cu(100) surface with a few vacancies visible. \textbf{c}, Conductance spectra acquired at the locations marked in panel \textbf{d}. \textbf{d}, Top: Vacancy dimer separated by a single Cl site. The coloured dots mark the locations of the point spectra in panel \textbf{c}. Centre: Conductance map at 3.3 V showing the bonding states. Bottom: Conductance map at 3.6 V showing the anti-bonding state. Note that this state lies within the conduction band for this structure and hence provides poor contrast \cite{Schuler2015}. \textbf{e}, Conductance spectra acquired at the locations marked in panel \textbf{f}. \textbf{f}, Top: Vacancy dimer separated by two Cl sites. The coloured dots mark the locations of the point spectra in panel \textbf{e}. Centre: Conductance map at 3.45 V showing the bonding states. Bottom: Conductance map at 3.53 V showing the anti-bonding state. All scale bars 1 nm.}
	\label{fig:Fig1_new}
\end{figure}

An overview image of the chlorine layer with several vacancies and typical d$I$/d$V$ spectra of the vacancy states are shown in Fig. \ref{fig:Fig1_new}a. The surface is characterised by a slowly varying density of states around the Fermi energy and a prominent band edge at ca. 3.5 eV (green line in Fig. \ref{fig:Fig1_new}a). The vacancy sites host an electronic state split off from this band edge (purple line).

We constructed vacancy dimers at different separations through lateral manipulation. Figures \ref{fig:Fig1_new}c-f show conductance spectra acquired on two vacancy dimers separated by one or two chlorine sites. The single peak of an individual vacancy splits into two components in both cases, with the splitting being larger for the smaller separation. Conductance maps acquired at the energies of the split peaks, presented in Figs. \ref{fig:Fig1_new}d and \ref{fig:Fig1_new}f, clearly show that the lower energy resonance is found between the vacancy sites, while the higher energy component is stronger on the outer edges (the high energy component of the strongly coupled dimer lies within the conduction band of the chlorine layer and thus gives poor contrast). These are clear signatures of the formation of bonding and anti-bonding combinations of the vacancy state wavefunctions \cite{Repp2005,Schuler2015}. Based on our data, we estimate the following hopping amplitudes: $t_{short} \approx 0.138$ eV, $t_{long} \approx 0.039$ eV and $t_{long}/t_{short} \approx 0.28$. These results establish our premise that the vacancy states may be used to construct interacting lattices. 

One of the simplest models in which topological states emerge is the SSH dimer chain, originally developed to describe the polyacetylene molecule and the soliton states within it \cite{heeger}. The topologically protected domain-wall states in the SSH model provide a condensed-matter realization of the domain-wall states in the Jackiw-Rebbi model of (1+1) dimensional Dirac equation with a mass kink \cite{jackiw1976}. The SSH Hamiltonian is given by:

\begin{equation}
	\mathscr{H} = -\sum_i t_{i} \hat{c}_i^{\dagger} \hat{c}_{i + 1}+ h.c.,
\end{equation}

\noindent where the hopping parameter alternates between $t_i = t_0 \pm \delta t$ for even/odd $i$. The system has a band gap determined by $|\delta t|$. The chain exists in two phases depending on the sign of $\delta t$ and distinguished by a topological index, the winding number. Physically, this means that the two phases are distinguished by the location of the strong bonds. Mid-gap states are expected on the boundary between these two phases where $\delta t$ changes sign. The characteristic localisation length of the domain-wall states in units of lattice constant is $\xi=|t_0-\delta t|/|\delta t|$. In our experiment, we expect strongly localised states as $\delta t \approx 0.7t_0$.

\begin{figure}
	\centering
		\includegraphics[width=0.90\textwidth]{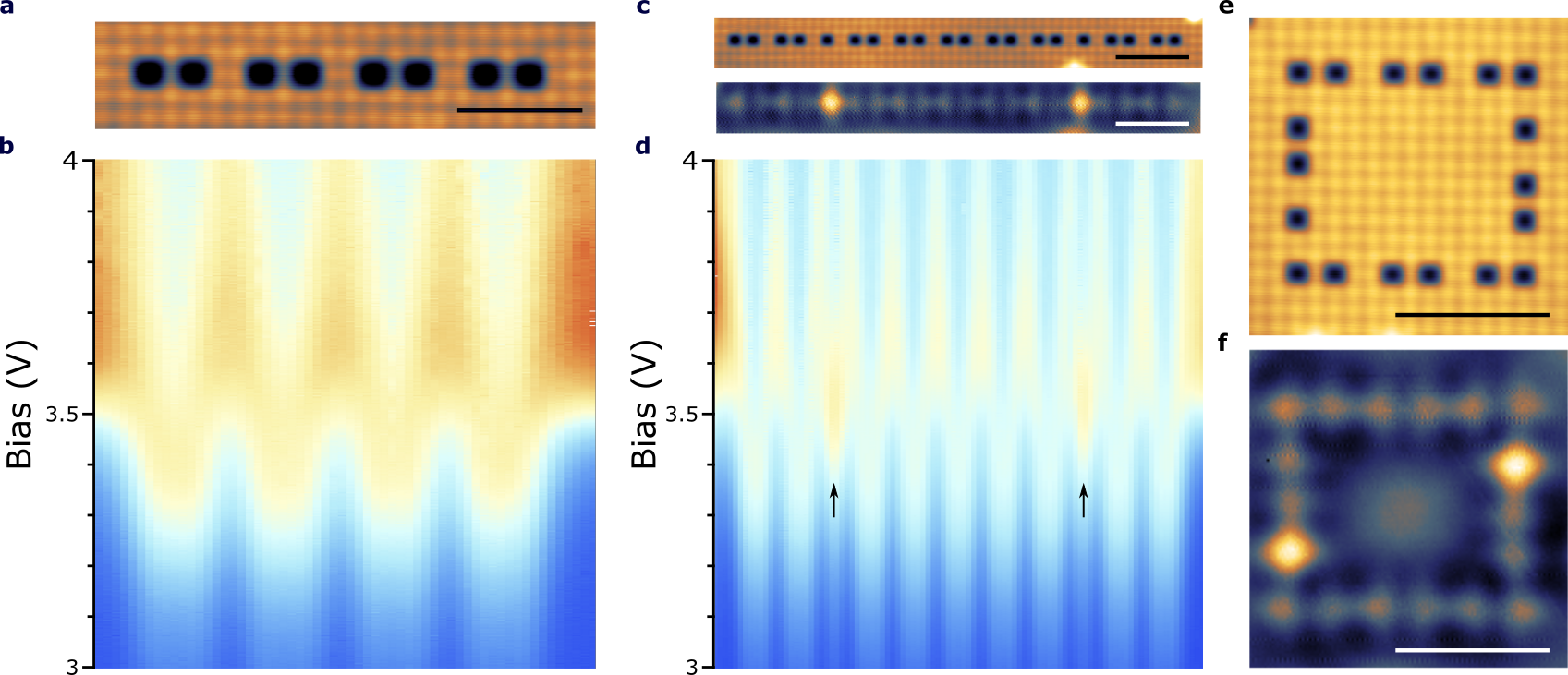}
	\caption{\textbf{Domain walls within dimer chains and rings.} \textbf{a}, Topography of a dimer chain assembled from vacancy site. Scale bar 2 nm. \textbf{b}, Stacked contour plot of point spectra taken along the axis of the chain in panel \textbf{a}. Red and blue correspond to high and low conductance, respectively. The states split into two bands with no states located within the gap. \textbf{c}, Top: Topography of a dimer chain with two domain walls separating sections with different topological properties. Bottom: Constant height conductance map acquired at 3.53 V (mid-gap) showing two prominent states at the locations of the domain walls. Scale bars 2 nm. \textbf{d}, Stacked contour plot of point spectra acquired along the axis of the chain in panel \textbf{c}. Two states appear at the mid-gap energy at the domain wall locations. \textbf{e}, Topography of a ring-like structure with two domain walls. Scale bar 3 nm. \textbf{f}, Conductance map acquired at 3.5 V showing mid-gap states at the domain wall locations.} 
	\label{fig:Fig3_new}
\end{figure}

The structure and a set of spectra for a short dimer chain can be found in Fig. \ref{fig:Fig3_new}a and b. Two sets of states can be seen to extend throughout the chain above and below the vacancy state energy. In analogy to SSH model, we refer to the region in between these two bands as the gap and to states within it as sub-gap states despite their location far above the Fermi energy. According to our estimation of the coupling constants, we expect a band gap of

\begin{equation}
	\Delta E_{min} = \sqrt{t_{short}^2 + t_{long}^2 - 2t_{short}t_{long}} \approx 0.18\; \mathrm{eV}
\end{equation}

\noindent which agrees with our experimental findings. No sub-gap states exist within this structure.

We constructed a chain containing two domain walls, see Fig. \ref{fig:Fig3_new}c. The lower panel of Fig. \ref{fig:Fig3_new}c shows a constant height conductance map of this structure acquired at 3.53 V. The two domain walls are clearly visible at this bias. We confirmed the existence of mid-gap modes through point spectroscopy. Fig. \ref{fig:Fig3_new}d shows a stacked contour plot of conductance spectra acquired along the chain. The bulk of the chain shows the same signatures as those shown in Fig. \ref{fig:Fig3_new}b. At the location of the domain walls however, a single resonance centred around the mid-gap energy emerges. These are the topologically protected mid-gap states predicted by the SSH model.  To show that the presence of topological states is independent of the chain form, we also studied a ring-like structure with two embedded phase boundaries. This structure can be found in Fig. \ref{fig:Fig3_new}e with the corresponding conductance map at the mid-gap energy in Fig. \ref{fig:Fig3_new}f. As expected, mid-gap states located on the domain wall sites are found.

To show that the concept of lattice engineering can be generalised to more complex systems, we constructed a Lieb lattice structure from chlorine vacancies. This is a line centred square structure with a three atom unit cell, resulting in a fermionic system with a lattice pseudospin of 1. The band structure in the infinite limit consists of a Dirac cone on the corners of the first Brillouin zone intersected by a flat band at the Dirac energy. Such flat bands are prone to electronic instabilities near half filling and have been suggested to yield magnetic or superconducting order. The presence of flat bands could also enhance the properties of superconducting materials by increasing the critical temperature \cite{Kopnin2011,julku2016}. Depending on the nature of the interaction in the lattice, topological states may also emerge \cite{weeks2010, goldman2011}. While Lieb lattices have been studied in optical lattices and ultra cold atomic gases \cite{mukherjee2015, taie2015}, no realisation using \textit{electronic} states exist to our knowledge.

\begin{figure}
	\centering
		\includegraphics[width=0.69\textwidth]{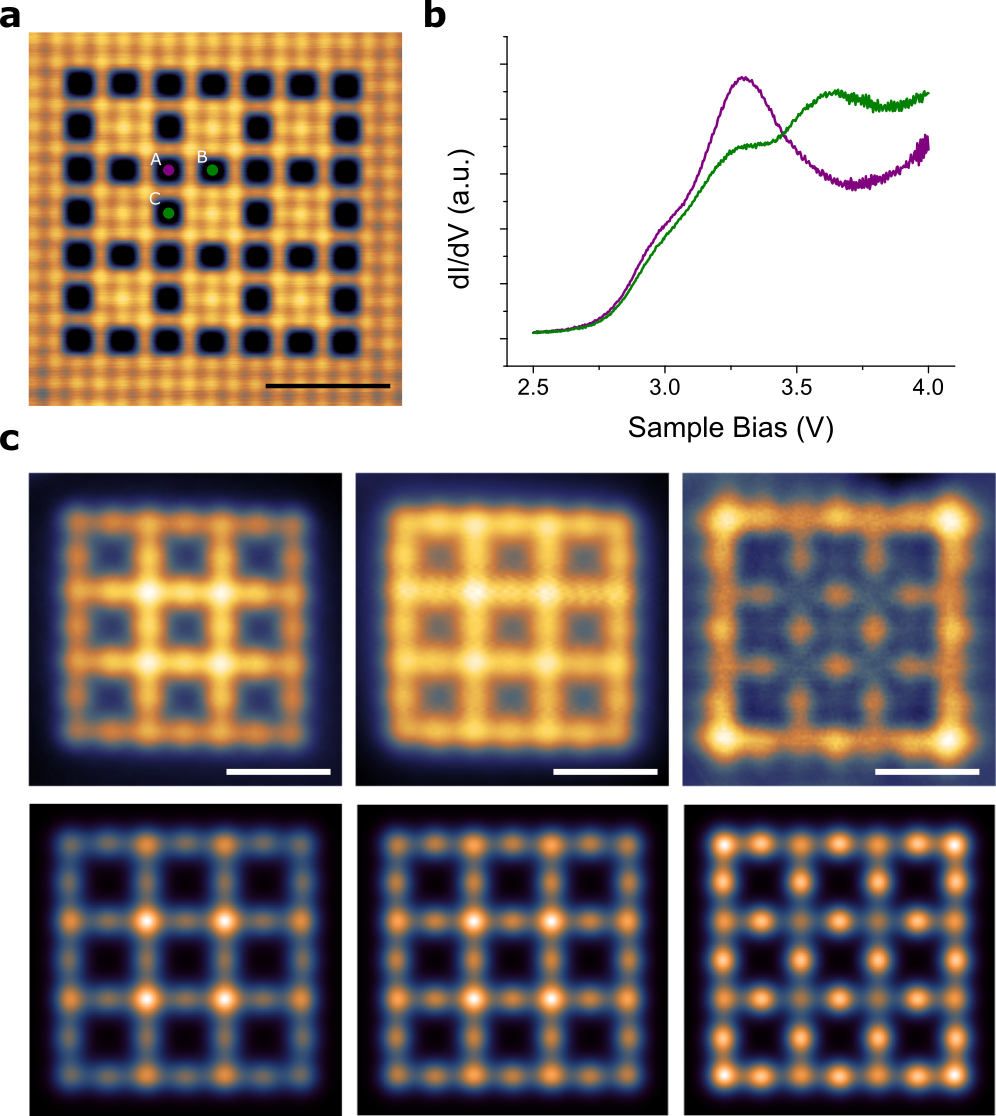}
	\caption{\textbf{Lieb lattice:} \textbf{a}, Topography of the Lieb lattice structure assembled from Cl vacancies. \textbf{b}, Point spectra acquires at the locations indicated in panel a. \textbf{c}, Conductance maps and tight binding simulations of the Lieb lattice. Bias valus (from left to right): 2.85 V, 3.15 V, and 3.5 V. The rightmost panel corresponds to the flat band. All scale bars 2 nm.} 
	\label{fig:Fig4_new}
\end{figure}

A small Lieb lattice can be seen in Fig. \ref{fig:Fig4_new}a with point spectra taken on the A, B and C sites of the three atom unit cell presented in panel b. Fig. \ref{fig:Fig4_new} shows constant height conductance maps acquired at different bias voltages. Between ~2.85 V and ~3.15 V the maps reveal an extended electronic state with higher intensity on the A sites. This finding is well reproduced by our tight binding simulation (see below). The signal distribution we observe here is characteristic of the dispersive bands in the Lieb lattice which converge to form the Dirac cone. The map taken at a bias of 3.5 V reveals a starkly different contrast: There now is nearly no signal on the A sub-lattice while the B and C sites show up more prominently. This is the hallmark of the the flat electron band for which the Lieb lattice is known.

We performed tight binding simulations of the Lieb lattice based on our experimental structure and estimates of the hopping amplitudes. As the next-nearest neighbour coupling is non-zero, the flat band will be slightly distorted. We estimate the next-nearest neighbour interactions by assuming that the coupling between vacancies depends exponentially on the separation distance to obtain a value of $t_{NNN} \approx 0.045 \; \mathrm{eV}$. The lower panel of Fig. \ref{fig:Fig4_new} shows the results of the tight binding simulation. A good agreement with the experiment is found using our coupling estimates of $t_{NN} \approx 0.14 \; \mathrm{eV}$ and $t_{NNN} \approx 0.045 \; \mathrm{eV}$, giving a ration of $t_{NNN}/t_{NN} \approx 0.33$  (see Supplementary Information for details). 

In summary, we have presented a general approach for producing tailor-made band structures through atom manipulation using the STM. We implemented two model systems with topological states and nearly flat electron bands in a precisely controlled environment. Our approach, combined with the possibilities of automating structure building at the atomic level, places a vast amount of relevant model systems within experimental reach. Additional customisability could come from adapting such assembly techniques to, for example, heavy metals with magnetic properties or significant spin-orbit coupling. The next major challenge is to identify systems in which the energies of the participating states are close to the Fermi level, or can be tuned through the relevant energy range, in order to observe effects such as the magnetic instability of the flat band or the fractional states of SSH chain. Further progress in this direction may allow atomic assemblies to produce intriguing physics similar to those seen in ultra-cold atomic gases while working entirely with electronic states. The same concepts laid out here may be applied to mesoscopic building blocks such as quantum dots in order to produce quantum materials with tailored properties.

\section*{Methods}
All sample preparations and experiments were carried out in an ultrahigh vacuum system with a base pressure of $\sim$10$^{-10}$ mbar. The (100)-terminated copper single crystal was cleaned by repeated cycles of Ne$^{+}$ sputtering at 1.5 kV, annealing to 600$^{\circ}$C. To prepare the chloride structure, anhydrous CuCl$_2$ was deposited from an effusion cell held at 300$^{\circ}$C onto the warm crystal ($T\approx$ 150 - 200$^{\circ}$C) for 180 seconds. The sample was held at the same temperature for 10 minutes following the deposition.

After the preparation, the sample was inserted into the low-temperature STM (Unisoku USM-1300) and all subsequent experiments were performed at $T=4.2$K. STM images were taken in the constant current mode. d$I$/d$V$ spectra were recorded by standard lock-in detection while sweeping the sample bias in an open feedback loop configuration, with a peak-to-peak bias modulation of 20 mV at a frequency of 709 Hz. Line spectra were acquired in constant height; the feedback loop was not closed at any point between the acquisition of the first and last spectra. Manipulation of the chlorine vacancies was carried out using a procedure adapted from Ref. \cite{Kalff2016}. The tip was placed above a Cl atom adjacent to a vacancy site at 0.5 V bias voltage and the current was increased to 1 to 2 $\mu$A with the feedback circuit engaged. The tip was then dragged towards the vacancy site at a speed of up to 250 pm/s until a sharp jump in the $z$-position of the tip was observed. This procedure lead to the Cl atom and the vacancy site exchanging positions with high fidelity.

\section*{Acknowledgements}
This research made use of the Aalto Nanomicroscopy Center (Aalto NMC) facilities and was supported by the European Research Council (ERC-2011-StG No. 278698 “PRECISE-NANO”), the Academy of Finland through its Centres of Excellence Program (projects no. 284594 and 284621), and the Aalto University Centre of Quantum Engineering.

\section*{Author Contributions}
All authors jointly conceived and planned the experiment. R.D. performed the measurements. R.D. and P.L. analysed the STM data. T.O. proposed the SSH dimer chain structure and provided the description of its physics. A.H. proposed the Lieb lattice model and performed the tight binding calculations for this structure. All authors jointly authored, commented, and corrected the manuscript.

\bibliography{bibliography}

\newpage
\section*{Supplementary Information}

\section*{Alternative Domain Wall}

\begin{figure}[b!]
	\centering
	\includegraphics[width=0.45\textwidth]{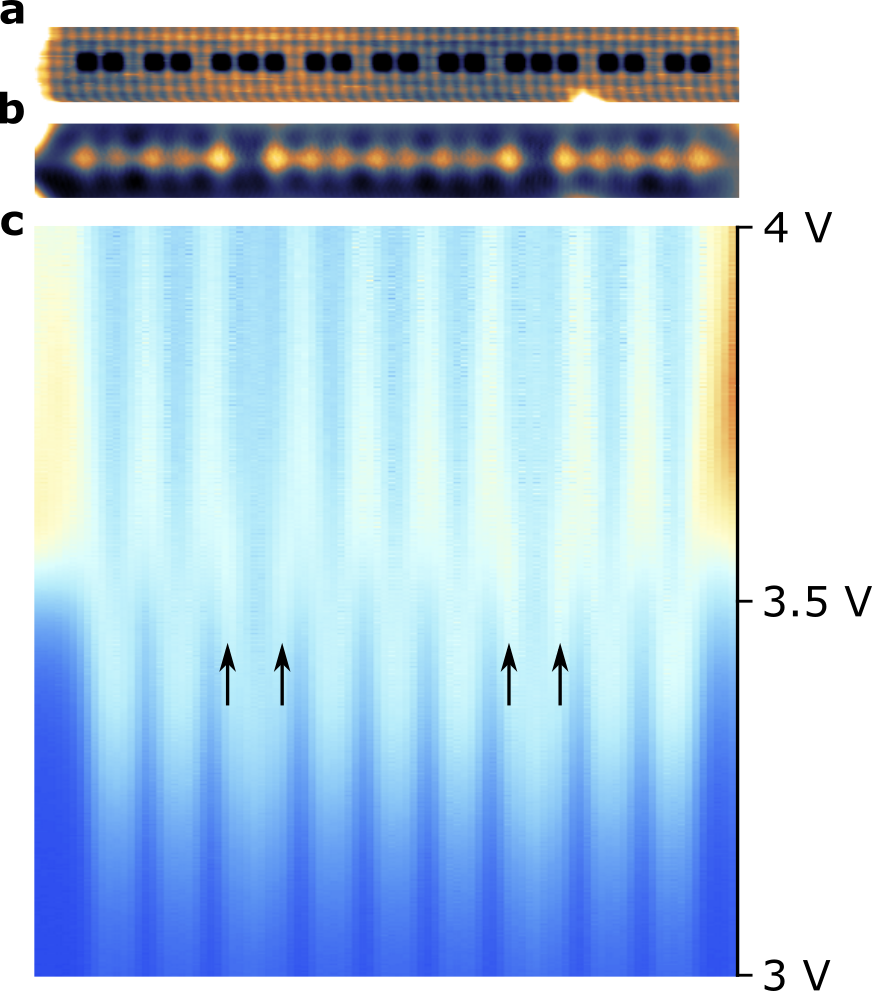}
	\caption{\textbf{Alernative domain wall:} \textbf{a}, Topography of the dimer chain with the alternative domain wall. \textbf{b}, Conductance map at 3.53 V (mid-gap energy). The domain wall state appears as an increased signal on the outer two sites of the vacancy trimer constituting the domain wall. \textbf{c}, Stacked contour plot of a series of conductance spectra acquired along the axis of the chain. The domain wall state is marked by the arrows.}
	\label{fig:SI_Fig1}
\end{figure}

The topological nature of the band structure in the dimer chains demands that there be a domain wall state associated with any transition from one of the phases to the other. This is demonstrated in the manuscript with the example of a domain wall in which sequence of the alternating coupling constants is changed by replacing a strong bond with a weak one. The arrangement results in a sequence of three lattice sites coupled by weak bonds as shown in Fig. 2 of the manuscript. Another domain wall structure exists whereby a weak bond in the chain is replaced by a strong one. This results in a sequence of three lattice sites coupled by strong bonds. We constructed a dimer chain with such a domain wall an investigated its band structure to confirm the topological nature of the domain wall states. The chain is shown in Figure \ref{fig:SI_Fig1}. 

The domain wall state is not localised to a single lattice site as it was the case with the structure shown in Fig. 2 of the manuscript. It instead appears on both of the outer sites of the domain wall trimer. As confirmed by our theoretical calculation of the LDOS, it is nevertheless a single electronic state which is spread across both lattice sites, thus fulfilling the topological requirement of a single mid-gap state per domain wall.

\section*{Estimating the hopping Parameters}

A background subtraction using a reference spectrum acquired on the chlorine layer far away from any vacancies resulted in highly asymmetric line shapes and was abandoned as a result. To estimate the peak positions, we instead used the half maximum of the rising flank on the first peak as a marker. This approach can obviously only be used to estimate the position of the first resonance in each spectrum. The hopping parameters in the dimers can then be calculated by subtracting the energy estimate for the bonding peak from that of a single vacancy.

In order to estimate the next-nearest neighbour coupling in the Lieb lattice, we assumed an exponential dependence of the hopping parameters with distance. This model was fitted to the available data set, consisting of the long and short dimers shown in the manuscript text and a diagonal dimer with two vacancies touching each other at the corners. 


\section*{Tight Binding Simulations of the Lieb Lattice}

\begin{figure*}
	\centering
	\includegraphics[width=.9\textwidth]{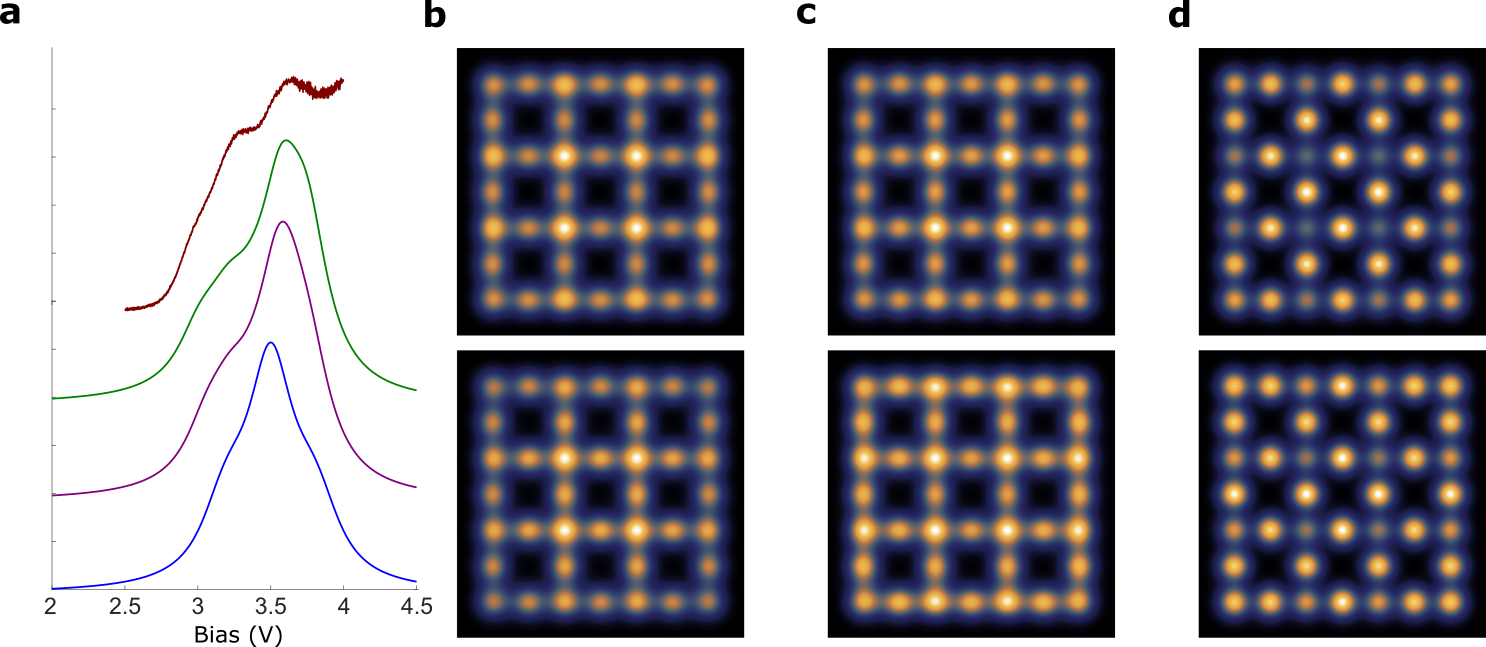}
	\caption{\textbf{Tight binding simulations with different coupling ratios:} \textbf{a}, LDOS on the B lattice site. An experimental curve is shown in red for reference. The simulated curves use the ratios $r_{NNN} = 0$ (blue), 0.34 (purple), and 0.5 (green). Curves are offset for clarity. \textbf{b} - \textbf{e}, Simulated conductance maps at 2.85 V, 3.15 V and 3.5 V (same as the experimental maps in the manuscript) with $r_{NNN} = 0$ (top) and $r_{NNN}$ = 0.33 (bottom)}
	\label{fig:SI_Fig3}
\end{figure*}

Based on our estimates of the coupling constants outlined above, we performed tight binding calculations on the Lieb lattice, as mentioned in the manuscript. Despite the rough estimate of the coupling constants, simulations with a nearest neighbour coupling of $t_{NN} \approx -0.138 \,$eV and a ratio of next-nearest neighbour to nearest neighbour coupling of $r_{NNN} = t_{NNN}/t_{NN} \approx 0.33$ reproduce the experimental results far better than those with higher or lower ratios $r_{NNN}$, see Fig. \ref{fig:SI_Fig3} for an illustration. The on-site energy was fixed to 3.5 V. The LDOS spectra and maps were calculated with an estimated broadening of 0.17 eV for all states.

The pronounced shoulder at 3.1 V in the spectra measured on the B lattice site is best reproduced for the simulated spectrum with $r_{NNN} = 0.33$. It is absent at lower ratios and splits into a double shoulder for higher $r_{NNN}$, which is not observed in the experiment. Some key features of the experimental conductance maps are also better reproduced using $r_{NNN} = 0.33$. In particular, the states are more confined towards he centre of the lattice at $E = 2.85$\,V, but more uniformly distributed at $E = 3.15$\,V. At the nearly flat band, simulations with $r_{NNN} = 0.33$ reproduce the increased signal at the centre of the edges of the structure. In addition, the signal at the corners of the structure is also increased. We speculate that the strong signal on the corner sites observed in the experiment is due to coupling to states in the conduction band of the chlorine layer.

\end{document}